# Transient birefringence of liquids induced by terahertz electric-field torque on permanent molecular dipoles

**Mohsen Sajadi\*, Martin Wolf and Tobias Kampfrath**
*Fritz-Haber-Institut der Max-Planck-Gesellschaft, Berlin, Germany*

\* E-mail: sajadi@fhi-berlin.mpg.de

**Microscopic understanding of low-frequency molecular motions in liquids has been a longstanding goal in soft-matter science. So far, such low-frequency motions have mostly been accessed indirectly by off-resonant optical pulses. A more direct approach would be to interrogate the dynamic structure of liquids with terahertz (THz) radiation. Here, we provide evidence that resonant excitation with intense THz pulses is capable of driving reorientational-librational modes of aprotic polar liquids through coupling to the permanent molecular dipole moments. We observe a hallmark of this enhanced coupling: a transient optical birefringence up to an order of magnitude higher than obtained with optical excitation. Our results open up the path to applications such as efficient molecular alignment and systematic study of the coupling of rotational motion to other collective motions in liquids.**

Low-frequency structural dynamics of liquids in the range from 0.1 to 10 THz (3 to 330 cm$^{-1}$) are believed to strongly contribute to the outcome of chemical processes[1,2,3,4,5]. The underlying molecular motions can be complex and include reorientations, vibrations and translations. To access and trigger rotational dynamics, one may take advantage of the torque

$$\boldsymbol{T} = (\boldsymbol{\mu}_0 + \boldsymbol{\mu}_{\text{ind}}) \times \boldsymbol{E} \qquad (1)$$

exerted on molecules by an external time ($t$)-dependent electric field $\boldsymbol{E}(t)$ (**Fig. 1a**). Coupling is mediated by (i) the permanent molecular dipole moment $\boldsymbol{\mu}_0$ (having constant modulus $\mu_0$) and (ii) the instantaneous dipole moment $\boldsymbol{\mu}_{\text{ind}}$ induced by polarizing the molecule's electron distribution.

For linearly polarized $\boldsymbol{E}$ and molecules with cylindrical symmetry, the torque due to $\boldsymbol{\mu}_{\text{ind}}$ scales with $\Delta\alpha E^2(t)$ where $\Delta\alpha$ quantifies the difference of the polarizability parallel and perpendicular to the molecular axis[6]. Since the square of $\boldsymbol{E}$ rectifies the rapidly oscillating light field, femtosecond laser pulses are routinely used to exert ultrafast torque of type $\boldsymbol{\mu}_{\text{ind}} \times \boldsymbol{E}$ on solvent molecules. In contrast, optical pulses yield a vanishing time-integrated torque of type $\boldsymbol{\mu}_0 \times \boldsymbol{E}$ because $\boldsymbol{E}$ changes the direction of the permanent dipole $\boldsymbol{\mu}_0$ very little over the only ~1 fs long optical half-cycle[7,8]. Therefore, to act on permanent dipole moments, we need to abandon the rapid field oscillations inherent to optical stimuli.

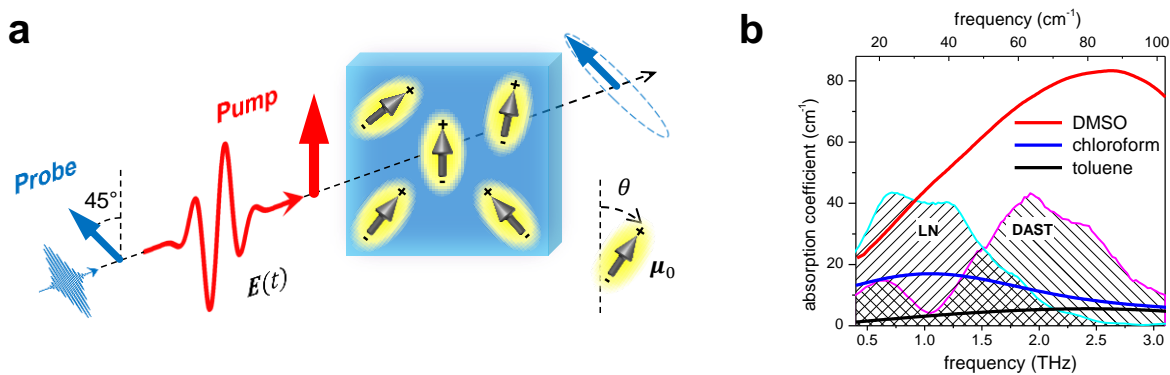

**Figure 1 | THz Kerr effect in dipolar liquids. a**, An intense THz or optical pump pulse induces birefringence in a polar liquid which is measured by an optical probe pulse that becomes elliptically polarized upon propagation through the medium. We study liquids with various values of the permanent molecular dipole moment $\mu_0 = |\boldsymbol{\mu}_0|$. **b**, Equilibrium THz absorption spectra of DMSO, chloroform and toluene. Amplitude spectra of THz pump pulses from two different sources (LN and DAST, see Methods) are shown by dashed areas.

Along theses lines, intense THz pulses were recently shown to align polar molecules in the gas phase through resonant excitation of their rotational transitions[9].

Here, we make use of THz electric fields to induce ultrafast $\boldsymbol{\mu}_0 \times \boldsymbol{E}$-like net torque on solvent molecules. We find that the resulting transient optical birefringence of polar liquids is enhanced by more than one order magnitude as compared to optical excitation. Our experimental observations are consistent with a simple model that provides insights into the interaction mechanism with the pump pulse, in part pointing to coupling of librational and reorientational modes. Thus, nonlinear THz spectroscopy exhibits large potential to provide new applications and new insights into low-frequency collective motions of liquids.

**Experiment.** A schematic of our experiment is shown in **Fig. 1a**. An intense, linearly polarized and phase-locked THz electromagnetic pulse (see Methods and amplitude spectra in **Fig. 1b**) is incident onto a polar liquid consisting of axially symmetric molecules. The resulting transient birefringence (THz Kerr effect, TKE)[10] is monitored by a time-delayed optical probe pulse whose polarization acquires an elliptical polarization. The degree of ellipticity scales with the difference[11,12]

$$\Delta n(t) \propto \int du\, \Delta\alpha\, f(u,t) P_2(u) \qquad (2)$$

between the liquid's optical refractive index perpendicular and parallel to the driving field $\boldsymbol{E}$. Here, $u = \cos\theta$ is the cosine of the angle between $\boldsymbol{E}$ and the molecular axis (**Fig. 1a**), $f(u,t)$ is its instantaneous distribution and $P_2(u)$ is $(3u^2-1)/2$. To directly compare the anisotropy induced by THz and optical excitation, we conduct the same measurements but with the THz pulse replaced by an optical pump pulse (optical Kerr effect, OKE). The instantaneous intensity of optical and THz pump pulse have approximately identical shape (see inset of **Fig. 2a**).

In our study, we focus on simple polar liquids, dimethyl sulfoxide (DMSO) and chloroform, for two reasons. First, their coupling to the incident THz field is predominantly mediated by the molecular inclination angle $\theta$ (**Fig. 1a**). Direct coupling to other degrees of freedom such as stretch coordinates is expected to be negligible at the frequencies <3 THz considered here. This notion is supported by the THz absorption spectra of the liquids which are shown in **Fig. 1b**. For chloroform, the amplitude spectrum of the THz pump (LN source) overlaps with both reorientational (at ~0.2 THz) and librational modes (~1 THz)[13]. For DMSO, the reorientational mode is at much lower frequencies (~10 GHz)[14], and THz pump spectra instead overlap with a librational mode[15].

Second, the liquids chosen here allow us to systematically compare the torque induced by coupling to induced electronic and to permanent dipoles because they exhibit distinctly different magnitude combinations of $\Delta\alpha$ and $\mu_0$: DMSO ($\mu_0 \approx 4.1$ D and $\Delta\alpha > 0$) and chloroform ($\mu_0 \approx 1.12$ D and $\Delta\alpha < 0$)[16]. To calibrate our comparative procedure, we perform OKE and TKE experiments on the nonpolar liquids toluene and cyclohexane ($\mu_0 \approx 0$).

**Transient birefringence.** The transient birefringence of three liquids following THz and optical excitation is shown in **Fig. 2**. For comparison, the squared THz pump field $E^2(t)$ and instantaneous optical pump intensity $I(t)$ are plotted in the inset of **Fig. 2a**. The signal amplitude is found to grow with the pump power, for both THz and optical excitation (Supplementary **Fig. S1**). Note that all TKE and OKE signals share two common features: (i) a sharp initial rise with a shape similar to the squared THz pump field $E^2(t)$ and optical intensity envelope $I(t)$, respectively (**Fig. 2a**), followed by (ii) a slower decay on a picosecond time scale.

Feature (i) is assigned to the response of the electronic subsystem of the sample[10,11,12]. This response is instantaneous because in our experiment, the excitation energies (>5 eV) of the electrons are much larger than the photon energies of the THz (<10 meV) and optical pump pulse (~1.5 eV).

Once pump and probe pulses do not overlap any more, the dynamics are dominated by the slower feature (ii) which is assigned to the relaxation of the nuclear degrees of freedom of the molecules. **Figure 2a** and the insets in **Figs. 2b** and **2c** reveal a remarkable observation for all three liquids: at pump-probe delays $t > 1$ ps, the dynamics are independent of the pump pulse used (THz or optical), apart from a global signal scaling factor. This finding shows that both THz and optical pump drive the same modes of the liquid. The mono-exponential birefringence decay of DMSO (time constant of ~6.4 ps) and the bi-exponential signal decay of chloroform (time constants ~0.4 and 2 ps) are in line with previous OKE studies[17,18], in which the slower time constants were assigned to reorientational relaxation.

To evaluate how efficiently these modes are excited by the THz and optical pump pulse, we normalize TKE and OKE signals to their respective peak value found around $t = 0$, as has already been done for the curves in the main panels of **Fig. 2**. As detailed in the Methods section, this procedure is tantamount to normalizing the signals to the pump intensity. Therefore, once pump-probe overlap is gone ($t > 1$ ps), normalized signal amplitudes approximately equal the relative strength with which THz and optical pump pulses drive the nuclear dynamics.

**Nonpolar vs polar liquids.** Interestingly, for toluene (**Fig. 2a**), identical normalized dynamic birefringence for THz and optical excitation are found at delays larger than the pump duration. Such agreement indicates that both THz and optical pump field couple to the rotational degrees of freedom with the same strength, consistent with our expectation: due to the relatively small permanent molecular dipole moment $\boldsymbol{\mu}_0$ of toluene, torque is dominated by the pump-induced moment $\boldsymbol{\mu}_{\text{ind}}$ (see Eq. (1)). As both the THz and optical pump photon energies are far off any electronic or vibrational resonance and since THz ($\boldsymbol{E}^2(t)$) and optical ($I(t)$) pump pulse have approximately identical shape (inset of **Fig. 2a**), both pulses exert identical normalized torques $\boldsymbol{\mu}_{\text{ind}} \times \boldsymbol{E}$ on molecules. This interpretation is confirmed by measurements on another nonpolar liquid, cyclohexane, where we also observe identical normalized dynamics following THz and optical excitation (see Supplementary **Figs. S2 and S3**).

Remarkably and in stark contrast to the nonpolar toluene, the normalized amplitude of the nuclear relaxation signal of the polar liquids DMSO (**Fig. 2b**) and chloroform (**Fig. 2c**) is seen to depend strongly on the pump frequency. While THz excitation of DMSO yields a ~10 times larger normalized birefringence signal than an optical pump (**Fig. 2b**), a signal reduction by a factor of ~3 is observed for chloroform (**Fig. 2c**). This observation indicates that in DMSO, the THz field couples more strongly to the rotational degrees of freedom than an optical pulse, whereas a reversed situation is found for chloroform. Note that this variation of coupling strength is strongly correlated with the material constants $\Delta\alpha$ and $\mu_0$. For $\mu_0 = 0$, identical birefringence is observed (**Fig. 2a**). In polar liquids ($\mu_0 > 0$), however, THz-induced birefringence is enhanced for $\Delta\alpha > 0$ (**Fig. 2b**) but reduced for $\Delta\alpha < 0$ (**Fig. 2c**) as compared to optical excitation. This trend is supported by data on three other liquids (see Supplementary

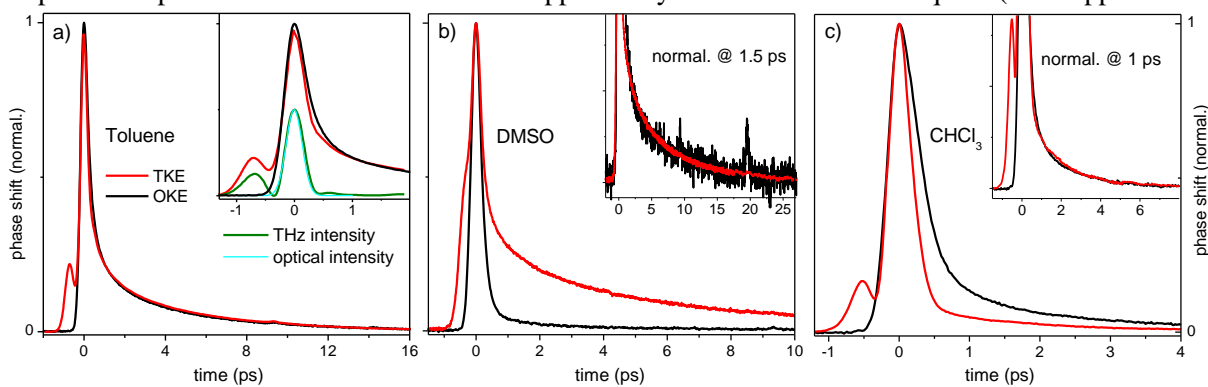

**Figure 2 | Transient optical birefringence of liquids following THz and optical excitation. a**, TKE (red line) and OKE (black line) signals of toluene. Signals are normalized to the initial peak signal where the electronic instantaneous electronic contribution is expected to dominate to the birefringence the birefringence signal. The inset shows the instantaneous intensity of the THz (green line) and optical pump pulse (cyan line). **b**, **c**, Same as panel **a**, but with data taken on DMSO (**b**) and chloroform (**c**), respectively. Insets in panels **b** and **c** show the same data but normalized to a delay of 1.5 ps and 1 ps, respectively.

**Fig. S4**).

**Model.** To obtain an interpretation of the measured birefringence dynamics $\Delta n(t)$ (**Fig. 2**), it is instructive to consider the impact of a THz or optical pump pulse on the angular distribution function $f(u, t)$ of the solvent molecules (**Fig. 3**). Since $\Delta n(t)$ is found to scale with the pump power (Supplementary **Fig. S1**), two interactions with the pump field are required. In equilibrium, the distribution function $f = f_0$ is isotropic and independent of the molecular inclination angle $\theta$ (**Fig. 3a**). When a $\delta$-like electric-field pulse is incident at time $t = t_1$, it exerts a torque through the permanent molecular dipole moment $\boldsymbol{\mu}_0$. The resulting perturbation $\Delta f_1(u, t)$ of the isotropic $f_0$ follows $P_1(u) = u = \cos\theta$ (**Fig. 3a**). Note that $\Delta f_1$ does not yet cause optical birefringence but is accompanied by a time-dependent dielectric polarization $\boldsymbol{P} = \chi^{\mu_0} * \boldsymbol{E}$. In this convolution, $\chi^{\mu_0}(t)$ is the $\boldsymbol{\mu}_0 \times \boldsymbol{E}$-type contribution to the dielectric susceptibility[19] that describes the temporal buildup and decay of $\boldsymbol{P}(t)$. Upon interaction with a second $\delta$-like field pulse at time $t = t_2$, part of $\Delta f_1$ is converted into a new distribution component whose shape follows $P_2(u) = (3u^2 - 1)/2$ (see **Fig. 3b** and Methods).

A $P_2$-like modification also arises from a single perturbation by the torque $\boldsymbol{\mu}_{\text{ind}} \times \boldsymbol{E} \propto \Delta\alpha E^2$ due to the induced electronic dipole moment $\boldsymbol{\mu}_{\text{ind}}$ (see **Fig. 3c**). The total $P_2$-type perturbation $\Delta f_2(u, t)$ due to the two torques causes transient optical birefringence that is measured in our experiment (Eq. (2)). By developing a simple but quite general model[20,21,22] for the dynamics of $f(u, t)$ (see Methods), we derive the transient optical birefringence,

$$\Delta n(t) \propto R_2 * [E \cdot (N\Delta\alpha E + 3\chi^{\mu_0} * E)]. \tag{3}$$

Here, $E(t)$ is the amplitude of the linearly polarized optical or THz pump field, and $N$ is the number of molecules per volume. Note that Eq. (3) reveals an analogy of the $\boldsymbol{\mu}_{\text{ind}}$- and $\boldsymbol{\mu}_0$-related coupling mechanisms: the first field interaction generates an effective electronic ($N\Delta\alpha E$) and orientational polarization ($\chi^{\mu_0} * E$) which, in turn, serves as a handle for the second field interaction to generate a $P_2$- like perturbation (square bracket in Eq. (3)). The decay of the resulting $P_2$-type change in the angular distribution function is captured by the response function $R_2(t)$, which is independent of the way the $P_2$-modification was generated.

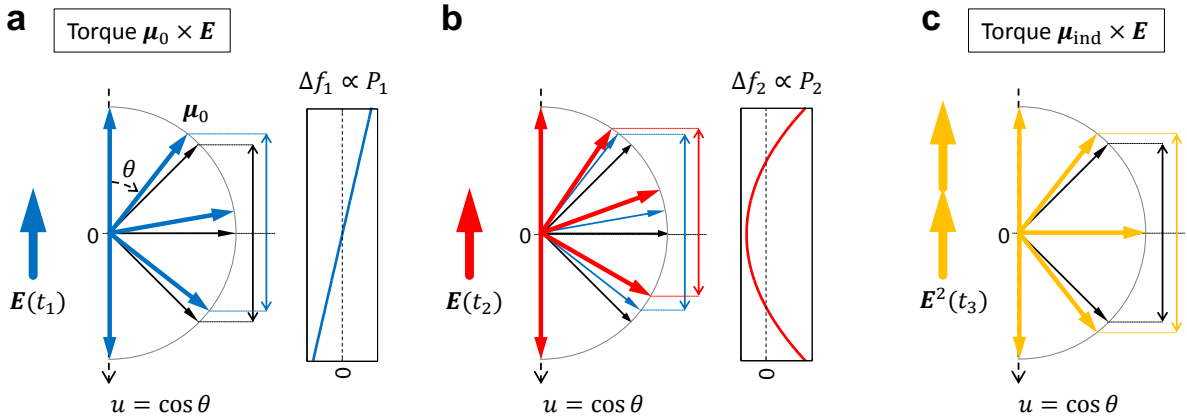

**Figure 3 | Model of transient optical birefringence. a**, In equilibrium, the orientation of permanent molecular dipoles is isotropic (black arrows), with a distribution function $f = f_0$ independent of the molecular inclination angle $\theta$. At time $t_1$, a field kick $E(t_1)$ exerts torque of type $\boldsymbol{\mu}_0 \times \boldsymbol{E}$ which orients molecules along $\boldsymbol{E}$, induces a $P_1$-type change $\Delta f_1$ in the distribution function and a dielectric polarization. Note that pairs $(\theta, 180° - \theta)$ of molecules are rotated rigidly. Therefore, the sum of the arrow lengths projected on the $u$-axis remains unchanged (see black and blue double arrows), and no optical birefringence is induced. **b**, At time $t_2 > t_1$, a second field kick triggers additional rotation, but in contrast to (**b**), dipoles in the lower hemisphere experience more torque than in the upper hemisphere. As a consequence, the change $\Delta f_2$ in the distribution function is $P_2$-like and accompanied by optical birefringence, as can be seen from the modified projected arrow lengths (blue and red double arrows) and from a quantitative model (Methods section). **c**, Torque of type $\boldsymbol{\mu}_{\text{ind}} \times \boldsymbol{E}$ at time $t_3$ scales with $E^2(t_3)$ and rotates each molecule to the closest pole (orange arrows), resulting in a $P_2$-type change in $f$ (see (**b**)) and optical birefringence (see black and orange double arrows).

We note that Eq. (3) is consistent with our central experimental findings: first, in all liquids studied, we observe identical relaxation dynamics for both optical and THz excitation (see **Fig. 2a** and insets in **Figs. 2b,c**). This agreement suggests that the picosecond decay of the optical birefringence is a manifestation of the $P_2$-relaxation function $R_2(t)$. Second, for $\mu_0 = 0$ and thus $\chi^{\mu_0} = 0$, no enhancement of birefringence is expected based on Eq. (3), in agreement with the experimental result (**Fig. 2a**). Third, Eq. (3) implies that for THz vs optical pumping, the normalized $\Delta n(t)$ is enhanced by a factor that scales with $1 + \chi^{\mu_0}/N\Delta\alpha \propto 1 + B\mu_0^2/\Delta\alpha$ where $B$ is a positive constant (see Methods). Thus, THz excitation should yield larger or smaller birefringence than optical excitation when polar liquids with, respectively, positive or negative $\Delta\alpha$ are used, in agreement with our experimental findings (**Figs. 2b,c**).

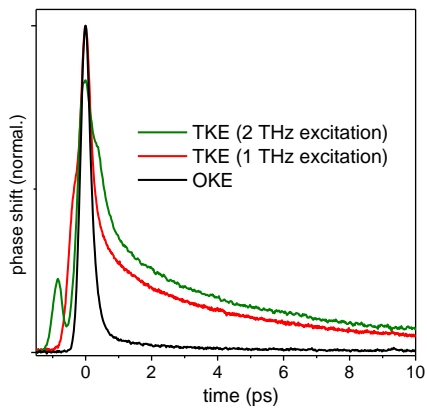

**Figure 4 | Impact of THz central frequency.** Normalized transient birefringence signals of DMSO following excitation with THz pump pulses centered at ~1 THz (red line) and ~2 THz (green line). The THz pump spectra are shown in **Fig. 1b**. For comparison, the signal induced by optical excitation is also shown (black line).

**Discussion.** According to Eq. (3), the THz-induced optical birefringence depends critically on $\chi^{\mu_0}$, the $\boldsymbol{\mu_0} \times \boldsymbol{E}$-related component of the dielectric susceptibility. Different modes of diffusive reorientation[19] (Debye modes) as well as hindered rotations[23] (librations) are known to contribute to $\chi^{\mu_0}$. The dielectric response of DMSO features a dominant Debye-type contribution[14], which peaks at ~10 GHz. Therefore, THz excitation is virtually off-resonant with respect to the Debye mode, and we expect reduced transient optical birefringence of DMSO when the THz pump frequency is increased (see Supplementary **Fig. S5**).

Note, however, the opposite trend is observed in the experiment, as seen in **Fig. 4**: when the center frequency of the THz pump is shifted from ~1 THz to ~2 THz, the birefringence signal increases notably. We assign this observation to resonant excitation of the librational mode of DMSO which is considered to be the origin of the broad absorption feature[15] at 2.5 THz (**Fig. 1b**). In our model of THz-induced birefringence, the first interaction with the pump field drives the libration (**Fig. 3a**), which, together with the second field interaction, induces an impulsive $P_2$-type perturbation. The resulting $P_2$-like distribution (**Fig. 3b**) decays according to rotational diffusion. Therefore, our results of **Fig. 4** provide a strong indication of coupling between librational and reorientational motions of DMSO.

In conclusion, we have conducted a systematic comparison of the transient optical birefringence in various liquids following optical vs THz excitation. Our experimental observations of increased/decreased birefringence are fully consistent with the notion that THz fields exert ultrafast torque on molecules due to their coupling to the permanent molecular dipoles. Our observation of resonant THz excitation suggests significant applications, for instance efficient molecular alignment of polar solutes and solvents, which may even enable coherent control of chemical reactions[24]. From a spectroscopic viewpoint, resonant and selective excitation of rotational motions can straightforwardly be extended to pairs of pump pulses[25,26,27,28]. Applied to hydrogen-bonded liquids such as water, such two-dimensional nonlinear THz spectroscopy of transient optical birefringence will provide fundamental insights into the coupling of modes associated with the collective rotational and translational motion of hydrogen-bond networks.

## Methods

**Experiment.** Intense THz fields at ~1 THz are generated by optical rectification in a 1.3 mol% MgO-doped stoichiometric LiNbO3 crystal (LN) with tilted pulse front technique[29,30]. The organic crystal 4-N,N-dimethylamino-4'-N'-methyl-stilbazolium tosylate (DAST) is used to generate intense THz pulses at ~2 THz[31]. The induced transient birefringence is measured by a temporally delayed probe pulses (2 nJ, 800 nm, 8 fs) derived from the seed laser oscillator[11]. Probe pulses are linearly polarized before the sample, encounter a transient birefringence induced by the co-propagating THz pulses, resulting in an elliptical polarization. The ellipticity is detected with a combination of a quarter-wave plate and a Wollaston prism which splits the incoming beam in two perpendicularly polarized beams with power $P_1$ and $P_2$. The normalized difference $(P_1 - P_2)/(P_1 + P_2)$ is twice the ellipticity and measured by two photodiodes as a function of temporal delay between THz pump and probe pulse.

To measure THz-field-induced optical birefringence of liquids, the choice of window materials is critical. Windows should be transparent at both THz (pump) and optical (probe) frequencies, optically isotropic, and their nonlinear THz responses should be small and short-lived. To fulfill all these criteria, we employ 200 nm thick SiN membranes as windows for a static cell (thickness of 100 μm). To make sure that accumulation of pump heat does not influence the results, we performed the TKE experiments also in a flow cell with the same SiN windows. We found no difference between static and flow cells in terms of both dynamics and amplitudes of the signals. The dried liquids were provided from Sigma Aldrich and used as received. To avoid wetting of the liquids, preparation of liquids and experiments were done under $N_2$ purging. Stationary THz absorption spectra of liquids were obtained with a broadband THz time-domain spectrometer based on a broadband spintronic THz emitter[32].

**Signal normalization.** Our comparative method is based on the fact that the electronic response of liquids is identical at THz and optical pump frequencies because the associated photon energies (~1.5 eV and ~10 meV, respectively) are much smaller than the electronic excitation energies (>5 eV) of the liquids studied here. By using a generic phenomenological model for the transient birefringence signal (Kerr effect), we show that nonpolar liquids (such as cyclohexane) exhibit identical Kerr response (see Supplementary **Section S2**). Therefore, since both our optical and THz pump pulse have approximately identical shape (see inset of **Fig. 2a**), almost identical normalized birefringence dynamics result following optical and THz excitation of nonpolar liquids (see **Fig. 2a** and Supplementary **Figs. S2 and S3**).

**Model details.** To develop a simple model that qualitatively describes the response of an ensemble of static rotors to an external electric field (optical and THz), we consider the dynamics of the angular distribution function $f(u,t)$. Here, $f(u,t)du$ quantifies the number of molecules having $u = \cos\theta$ in the interval $[u, u+du]$ at time $t$ where $\theta$ is the angle between the molecular dipole and the direction of the applied electric field (see **Fig. 3**). In equilibrium, $f$ equals $f_0 = N/2$, proportional to the particle density $N$ yet independent of $\theta$ (**Fig. 3a**).

In the rotational diffusion model, the equation of motion of $f(u,t)$ is given by[33]

$$(\partial_t + \hat{O})f = C\partial_u[(T^{\mu_0} + T^{\Delta\alpha})f] \tag{4}$$

where the operator $\hat{O}$ captures the dynamics of the system in the absence of external perturbations. The right-hand side of Eq. (1) describes the action of the external linearly polarized field with amplitude $E(t)$ through the torque $T^{\mu_0} \propto \sin^2\theta\, \mu_0 E$ (mediated by the permanent dipole moment $\boldsymbol{\mu}_0$) and the torque $T^{\Delta\alpha} \propto \Delta\alpha E^2 \sin\theta \cos\theta$ (mediated by the field-induced electronic dipole moment $\boldsymbol{\mu}_{\text{ind}}$), consistent with Eq. (1).

Note that in the rotational diffusion model, $\hat{O}$ is proportional to the Laplace operator, which is only capable of describing a random-walk-like relaxation of $f$ back to the equilibrium distribution $f_0$. We assume that additional rotational modes such as librations can be taken into account by an appropriately modified operator $\hat{O}$ (see Ref. 34). The constant $C$ in Eq. (1) is eventually fixed by comparing the final result to the known solution for a static electric field[35].

Since in our experiment, the transient optical birefringence (Eq. (2)) has been found to scale quadratically with the incident pump field at both optical and THz excitation, we need to solve Eq. (4) up to second order in the applied electric field. Thus, the general solution has the structure $f = f_0 + \Delta f_1 + \Delta f_2$ where $\Delta f_1$ and $\Delta f_2$, respectively, are contributions linear and quadratic in $E(t)$. By virtue of Eq. (4), $\Delta f_1$ and $\Delta f_2$ are found to obey the following differential equations[21],

$$(\partial_t + \hat{O})\Delta f_1 = C\partial_u(T^{\mu_0} f_0) \quad (5)$$

$$(\partial_t + \hat{O})\Delta f_2 = C\partial_u(T^{\mu_0}\Delta f_1 + T^{\Delta\alpha} f_0) \quad (6)$$

As $\Delta f_1$ and $\Delta f_2$ result from one and two interactions with the field of the pump pulse, it is often instructive to consider $\Delta f_1$ as the response to one $\delta$-like perturbation (**Fig. 3a**) and $\Delta f_2$ as the response to two subsequent $\delta$-like perturbations (**Fig. 3b**). From these impulse responses, the general linear and quadratic response can easily be determined.

As seen from the right-hand-side of Eq. (5), $\Delta f_1$ arises from the perturbation $C\partial_u(T^{\mu_0} f^0)$, which is proportional to the first-order Legendre polynomial $P_1(u) = u = \cos\theta$. This $P_1$-type perturbation suggests the resulting response $\Delta f_1$ has also approximately $P_1$-like characteristics, $\Delta f_1(u,t) \propto P_1(u)$. Indeed, this assumption has been shown to be precisely valid for rotational diffusion[33,35], and it can be further bolstered by the schematic of **Fig. 3a**: in case of a $\delta$-like field pulse, a $\delta$-like torque is exerted which instantaneously increases the mean angular velocity $\dot{\theta}$ of a molecule to a value proportional to the time-integrated torque, that is, $\sin\theta$. Therefore, as seen in **Fig. 3a**, population is shifted from the south pole ($\theta = 180°$) to the north pole ($\theta = 0°$), whereas it remains constant at the equator ($\theta = 90°$), consistent with a $\cos\theta$-type distribution change.

Note that the alignment of the molecular dipoles implies a polarization $P(t) \propto \mu_0 \int du\, P_1(u)\Delta f^1(u,t)$ along the field direction. This polarization is usually expressed by the convolution $P = \chi^{\mu_0} * E$ where $\chi^{\mu_0}$ is the contribution of the permanent dipoles to the familiar dielectric susceptibility, which can be measured by microwave and THz absorption spectroscopy[19]. The $\chi^{\mu_0}(t)$ captures the full dynamics of the $P_1$-component of $\Delta f_1$, from the buildup (e.g. due to initial directed rotation) to the final decay (e.g. due to rotational diffusion). Thus, the change in the angular distribution function arising from a first interaction with the field of the pump pulse can generally be written as

$$\Delta f_1(u,t) \propto P_1(u) \cdot (\chi^{\mu_0} * E)(t) \quad (7)$$

According to Eq. (6), the second-order response arises from the term $C\partial_u(T^{\mu_0}\Delta f_1 + T^{\Delta\alpha} f_0)$, which is proportional to the second-order Legendre polynomial $P_2(u) = 3u^2 - 1$. Therefore and analogous to the linear case, we assume this perturbation leads to a $P_2$-like change in the distribution function, both for the $\boldsymbol{\mu}_0 \times \boldsymbol{E}$- (**Fig. 3b**) and $\boldsymbol{\mu}_{\text{ind}} \times \boldsymbol{E}$-type torque (**Fig. 3c**). The temporal dynamics of $\Delta f_2$ are described by the response function $R_2$ which captures the build-up and (possibly oscillatory) decay of an impulsively induced $P_2$-distribution. The $R_2(t)$ can, in principle, be measured by the ultrafast optical Kerr effect because optical fields induce exclusively and impulsively the $P_2$-like perturbation $C\partial_u(T^{\Delta\alpha} f_0)$.

By evaluating the right-hand side of Eq. (6) by means of Eq. (7), subsequent convolution with $R_2$ yields

$$\Delta f_2(u,t) \propto P_2(u) \cdot \{R_2 * [E \cdot N\Delta\alpha E + 3E \cdot (\chi^{\mu_0} * E)]\}(t) \quad (8)$$

This result is consistent with the rotational diffusion model, which delivers mono-exponentially decaying step functions $\chi^{\mu_0}(t) = \chi^{\mu_0}_{\text{DC}}\Theta(t)\exp(-2Dt)$ and $R_2(t) \propto \Theta(t)\exp(-6Dt)$ for the two response functions[33]. Here, $D$ is the diffusion constant, $\Theta(t)$ is the Heaviside step function and $\chi^{\mu_0}_{\text{DC}} = N\mu_0^2/3k_\text{B}T$ is the susceptibility for a static electric field with $k_\text{B}T$ being the thermal energy. Finally, by projecting $\Delta f_2(u,t)$ [Eq. (8)] onto $P_2(u)$ [see Eq. (2)], we obtain the optical birefringence $\Delta n(t)$ [see Eq. (3)].